\documentclass[%
 reprint,
superscriptaddress,
 amsmath,amssymb,
 aps,
pra,
]{revtex4-1}

\usepackage{latexsym,amsmath,amssymb}
\usepackage{braket} 
\usepackage{ftnxtra}
\usepackage{fnpos}
\usepackage{tikz}
\usepackage{color}
\usepackage{footnote}
\newcommand{\xinhui}[1]{\textcolor{black}{#1}}

\newcommand{\nn}{\nonumber}

\usepackage{amsthm} 
\usepackage{appendix}

\newtheorem{theorem}{Theorem}

\newtheorem{definition}{Definition}[theorem]

\usepackage[T1]{fontenc} 

\usepackage{hyperref}
\hypersetup
{
colorlinks=true,
linkcolor=blue,
filecolor=blue,
urlcolor=blue,
citecolor=cyan,
}

\usepackage{graphicx}

\newcommand{\ba}{\begin{eqnarray}}
\newcommand{\ea}{\end{eqnarray}}
\newcommand{\ban}{\begin{eqnarray*}}
\newcommand{\ean}{\end{eqnarray*}}

\begin{document}
\title{
Self-testing of different entanglement resources via fixed measurement settings}
\author{Xinhui Li}
 \affiliation{National Laboratory of Solid State Microstructures, School of Physics, and Collaborative Innovation Center of Advanced Microstructure, Nanjing University, Nanjing, 210093, China}
\author{Yukun Wang}
\email{wykun06@gmail.com}
\affiliation{Beijing Key Laboratory of Petroleum Data Mining, China University of Petroleum, Beijing, 102249, China}
\affiliation{State Key Laboratory of Cryptology, P.O. Box 5159, Beijing, 100878, China}

\author{Yunguang Han}
\email{hanyunguang@nuaa.edu.cn}
\affiliation{College of Computer Science and Technology, Nanjing University of Aeronautics and Astronautics, Nanjing, 211106, China}

\author{Shi-Ning Zhu}
 \affiliation{National Laboratory of Solid State Microstructures, School of Physics, and Collaborative Innovation Center of Advanced Microstructure, Nanjing University, Nanjing, 210093, China}

\begin{abstract}

Self-testing, which refers to device independent characterization of the state and the measurement, enables the security of quantum information processing task  certified independently of the operation performed inside the devices. 
Quantum states lie in the core of self-testing as key resources. 
However, for the different entangled states, usually different measurement settings should be taken in self-testing recipes. This may lead to the redundancy of measurement resources. In this work, we use fixed two-binary measurements and answer the question that what states can be self-tested  with the same settings. 
By investigating the structure of generalized tilted-CHSH Bell operators with sum of squares decomposition method, we show that a family of two-qubit entangled states can be self-tested by the same measurement settings. The robustness analysis indicates that our scheme is feasible for practical experiment instrument. Moreover, our results can be applied to various quantum information processing tasks. 



\end{abstract}

\maketitle

\section{Introduction}
Bell nonlocality \cite{CHSH1969,Nonlocality2014} is central to the understanding of quantum physics. With the advent of quantum information, Bell nonlocality has been studied as a resource and applied to various quantum information processing tasks, such as quantum key distribution \cite{DIQKD2007,DIQKD2014}, randomness expansion \cite{DIQRNG2010,DIQRNG2018} and 
entanglement witness \cite{DIEW2013,DIEW2018}. 

Moreover, if we assume quantum mechanics to be the underlying theory, it is shown that certain extremal quantum correlations uniquely identify the state and measurements under consideration, a phenomenon known as  self-testing \cite{Mayers2004,2012Scarani}. It is a concept of device independence  whose conclusion verdict relies only on the observed statistics of measurement outcomes under the sole assumptions of no-signaling and the validity of quantum theory \cite{Supic2020}. In 1990s, Popescu and Rohrlich et al. pointed out that the maximal violation of the Clauser--Horne--Shimony--Holt (CHSH) Bell inequality identifies uniquely the maximally entangled state of two qubit \cite{Tsirelson,MI2011Gisin}. In the last decades, self-testing has received substantial attention. The scenarios for bipartite and multipartite entangled states were presented in Refs. \cite{2017Coladangelo,2015Bamps,2017Coladangelo,2016Wang,Wu1, 2014Wu,McKague2,2018Li,2020Li}. 
The robustness analysis to small 
deviations from the ideal case for self-testing these quantum states
and measurements were presented in Refs. \cite{2016Kaniewski,2019LWH,Bancal,2019Coopmans},
which made self-testing more practical. Beyond these works focusing on the single copy states, the parallel self-testing of tensor product states have recently been studied. The first parallel self-testing protocol was proposed for 2 EPR pairs in \cite{2016Wu}. The result was subsequently generalised for arbitrary $n$, via parallel repetition
of the CHSH game in \cite{Coladangelo} and
via parallel repetition of the magic square game
in \cite{Coudron}. Self-testing of $n$ EPR pairs via parallel
repetition of the Mayers-Yao self-test is given in \cite{2016McKague}.

In the most previous scenario, one measurement setting is always competent to self-test one 
target state up to local unitaries. \xinhui{For example, the tilted-CHSH inequality can self-test  two-qubit pure states $\ket{\psi(\theta)}=\cos\theta\ket{00}+\sin\theta\ket{11}$ with corresponding measurements settings $\{\sigma_z,\sigma_x\}\otimes\{\cos\mu \sigma_
z+\sin\mu \sigma_x,\cos\mu \sigma_z-\sin\mu \sigma_x\}$, meanwhile $\mu$ is uniquely determined by $\theta$.} However, the tasks of quantum information processing may involve multiple states with different entanglement degree \cite{2020Wagner}. The whole self-testing of quantum states results in an increased consumption of the measurement resource, thus strike the feasibility of practical realization. Therefore, self-testing protocol with high practical performance is meaningful and necessary.
In this work, we focus on this goal and provide a device independent scheme that certify a series of quantum states with reduced measurement resource.  \xinhui{ Our results show that the generalized tilted-CHSH operators allow the optimal measurements for one
party could rotate on Pauli $x-z$ plane. Multiple different target states can be self-tested via a common measurement settings by choosing proper generalized tilted-CHSH operator. 
Hence, by utilizing a set of Bell inequalities, we can self-test two-qubit states with different entanglement degree only based on two binary measurements per party. Thus our scheme simplifies the measurement instruments and leads to less consumption of measurement resources.} Besides, our scheme demonstrates satisfactory robustness in tolerance of noise. Further, our scheme can serve for various quantum information processing tasks with low measurement resources cost, meanwhile provides secure certification of the device used in the task. The paper is structured as follows: In Sec. \ref{sec:Preliminaries}, we give a brief description about the underlying 
model and key definitions of our work. 
In Sec. \ref{sec:tiltedCHSH}, we propose a scheme that self-tests different two-qubit entangled states via the same measurements using generalized tilted-CHSH inequality. During this study, we develop a family of self-testing criteria beyond the standard tilted-CHSH inequality and prove these criteria using the technique of sum-of-squares (SOS) decomposition. In Sec. \ref{Sec:Rob}, the robustness analysis is illustrated through an example by the swap method and semidefinite programming (SDP). In Sec. \ref{sec:Apl}, the applications of our results on quantum information processing tasks of device independent quantum key distribution, private query and randomness generation are presented.
In Sec.\ref{sec:Conclusion}, we summarize the results and discuss the future research.

\section{Self-testing different entangled states via two binary measurements }\label{sec:tiltedCHSH}

\subsection{Self-testings}\label{sec:Preliminaries}

Consider the simplest scenario of two noncommunicating parties, Alice and Bob. Each has access to a black box with inputs denoted respectively by $x,y\in \{0,1\}$ and outputs $a,b\in \{+1,-1\}$. One could model these boxes with an underlying state $\ket{\psi}_{AB}$ and measurement projectors $\left\{M_x^a\right\}_{x,a}$ and $\left\{M_y^b\right\}_{y,b}$, which commute for different parties. The state can be taken pure and the measurements can be taken projective without loss of generality, because the dimension of the Hilbert space is not fixed and the possible purification and auxiliary systems can be given to any of the parties. After sufficiently many repetitions of the experiment one can estimate the joint conditional statistics, as known as the behavior, $p(a,b|x,y) = \bra{\psi} M_x^aM_y^b \ket{\psi}$. 
\xinhui{Self-testing refers to a device-independent certification way where the nontrivial information on the state and the measurements is uniquely certified by the observed behavior $p(ab|xy)$, without assumptions on the underlying degrees of freedom. Usually, self-testing can be defined formally in the following way.}

\begin{definition}\label{Def1}
We say that the correlations $p(a,b|x,y)$ allow for self-testing if for every quantum behavior
$(\ket{\psi}$, $\{M_x^a,M_y^b\})$ compatible with $p(a,b|x,y)$ there exists a local isometry $\Phi=\Phi_A\otimes\Phi_B$ such that 
\begin{align}\label{eq:self-testing definiton}
\Phi\ket{\psi}_{AB}\ket{00}_{A'B'}
&=\ket{junk}_{AB}\otimes\ket{\overline{\psi}}_{A'B'}\\
\Phi(M^a_x \ket{\psi}_{AB}\ket{00}_{A'B'})
&=\ket{junk}_{AB}\otimes \overline{M^a_x}\ket{\overline{\psi}}_{A'B'},\nn
\end{align}
where $|00\rangle_{A'B'}$ is the trusted auxiliary qubits attached by Alice and
Bob locally into their systems, $(\ket{\overline{\psi}},\{\overline{M_x^a},\overline{M_y^b}\})$ are the target system \cite{2012Scarani}.
\end{definition}
\xinhui{That is the correlations $p(a,b|x,y)$ predicted by quantum theory could determine uniquely the state 
and the measurements, up to a local isometry}.


\subsection{Self-testing of entangled two-qubit states with generalized tilted-CHSH inequality }\label{sec:tiltedCHSH}

\xinhui{In this section, we show that different pure entangled two-qubit states can be self-tested via fixed measurement setting with generalized tilted-CHSH inequality.  The candidate target states we considered are $\{\ket{\overline{\psi}_i}\}$, with
\begin{align}
    \ket{\overline{\psi}_i}&=\cos\theta_i\ket{00}+\sin\theta_i\ket{11}\label{eq:St2}
\end{align}
where $\theta_i\in(0,\frac{\pi}{4}]$. It is already proved that pure entangled two-qubit state can be self-tested using standard tilted-CHSH inequality \cite{2015Bamps,2019Coopmans}. In the standard scheme, one measurement setting is required for self-testing one target state, which results in an increased consumption of the measurement resource. Utilizing the property of generalized generalized tilted-CHSH inequality, we show that all these entangled states can be self-tested  with the given fixed measurements, thus simplifies the measurement instruments.} We have the following theorem.

\begin{theorem}\label{Theorem1}
 The family of entangled two-qubit states in Eq. \eqref{eq:St2}  can be self-tested  by  the same  quantum measurement settings in Eq. \eqref{eq:M} with fixed angle $\mu$. The fact of the self-testing result comes from the maximum quantum violation of generalized tilted-CHSH inequalities  in Eq. \eqref{eq:T2}.
\end{theorem}

The measurements in our scheme are chosen as, 
 \begin{align}\label{eq:M}
 A_0&=\sigma_z,\;\;B_0=\cos\mu\sigma_z+\sin\mu\sigma_x\nn\\
 A_1&=\sigma_x,\;\;B_1=\cos\mu\sigma_z-\sin\mu\sigma_x
 \end{align} with the fixed angle $\mu\in(0,\frac{\pi}{4}]$. 
 
The key idea of our self-testing scheme is that for a given $\mu$ in the unit measurement settings, 
a family of Bell inequalities can be maximally violated by different entangled pairs respectively at the same time. Once the form of the target source is confirmed, the Bell inequality which achieve their self-testing are determined based on the observed statistics, $p(a,b|x,y)$. 
More precisely, the Bell inequalities have the form conditional on the input $i\in\{0,1,2,..,n\}$ as
\begin{align}
\mathcal{B}_{[\alpha_i,\beta_i]}=&\beta_i A_0+\alpha_i (A_0B_0+A_0B_1)+A_1B_0-A_1B_1, \label{eq:T2}
\end{align}
named generalized tilted CHSH inequality \cite{Acin2012}, where $\alpha_i\geq 1$. The  maximal classical and quantum bounds  are  
$C_{[\alpha_i,\beta_i]}=2\alpha_i+\beta_i$ and $\eta_{[\alpha_i,\beta_i]}=\sqrt{(4+\beta_i^2)(1+\alpha_i^2)}$, respectively. \xinhui{It is already proved both theoretically and numerically that pure entangled two-qubit state can be self-tested using standard tilted-CHSH inequality with $\alpha=1$ in Eq. \eqref{eq:T2}  \cite{2015Bamps,2019Coopmans}. However, whether generalized tilted-CHSH inequality can be used in self-testing is still unknown. } 




\begin{figure}[ht!]
\centering
\includegraphics[scale=0.5]{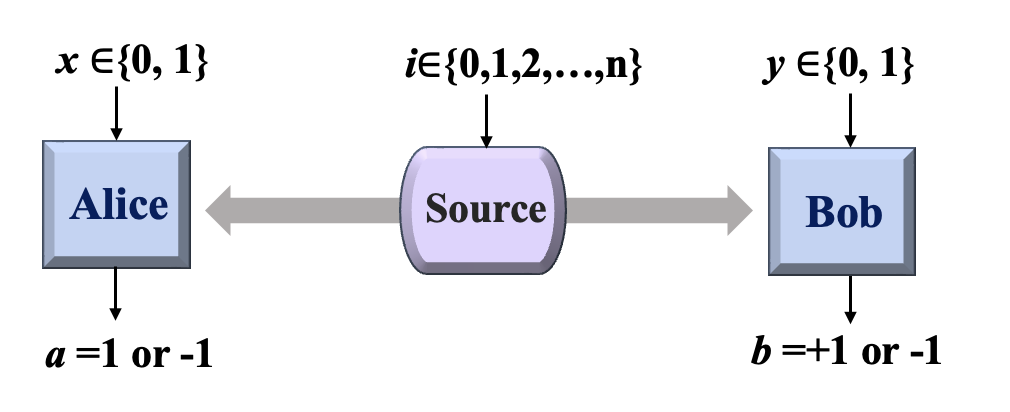}
\caption{Self-testing recipe. The fixed untrusted measurements setting are able to test different input states. For a given a target state $\ket{\overline{\psi}_i}$, two observers randomly chose their measurement $x$ for Alice, $y$ for Bob, and collect  outcomes $a$ and $b$ to construct the observation of $\mathcal{B}_{[\alpha_i,\beta_i]}$.  }\label{Fig:scheme}
\end{figure}

We claim that the maximal quantum violation in Eq. \eqref{eq:T2} uniquely certify the corresponding entangled pairs in Eq. \eqref{eq:St2} and measurements in Eq. \eqref{eq:M} with $\sin2\theta_i=\sqrt{\frac{4-\alpha_i^2\beta_i^2}{4+\beta_i^2}}$ and $\tan\mu=\frac{\sin2\theta_i}{\alpha_i}$. \xinhui{Thus the family of pure entangled two-qubit states are self-tested with the given $\alpha$, $\beta$ and the fixed two measurement settings $\mu$ using generalized tilted-CHSH inequality. The self-testing recipe of our scheme is shown in Fig. \ref{Fig:scheme}.  In the following, we give the detailed proof of Theorem \ref{Theorem1}. } 

\textbf{Proof.} \xinhui{The proof of Theorem \ref{Theorem1} is divided into two steps. First, we  give two types of SOS decompositions for generalized tilted-CHSH operator $\mathcal{B}_{[\alpha,\beta]}$ (see Appendix for details). 
Moreover, these SOS decompositions  establish algebraic relations that are necessarily satisfied by any quantum state and observables yielding maximal violation of the generalized tilted-CHSH inequality. Then these algebraic relations are used in the isometry map to  provide the self-testing of any partially entangled two-qubit state.}

\textit{a. SOS decompositions for generalized tilted-CHSH inequalities.}
The generalized tilted-CHSH inequalities
$\mathcal{B}_{[\alpha,\beta]}$ have the maximum quantum violation value
 $\eta_{[\alpha,\beta]}$. 
This implies that the operator $
\widehat{\mathcal{B}}=\eta_{[\alpha,\beta]}\mathbb{I}-\mathcal{B}_{[\alpha,\beta]}
$
is positive semidefinite for all possible quantum states and measurement
operators $A_x$ and $B_y$. This can be proven by providing a set of operators $\{P_i\}$
which are polynomial functions of $A_x$ and $B_y$
such that $
\widehat{\mathcal{B}}_{[\alpha,\beta]}=\sum_i P^\dagger_i P_i
$,
holds for any set of measurement operators satisfying the
algebraic properties $A^2_x=\mathbb{I}$, $B^2_y=\mathbb{I}$ and
$[A_x,B_y]=0$.

For convenience, we define three classes CHSH operators:
\begin{align}
S_0&= A_0(B_0-B_1)+\frac{1}{\alpha}A_1(B_0+B_1),\nn\\
S_1&=\frac{1}{\alpha}A_0(B_0+B_1)-A_1(B_0-B_1),\\
S_2&=A_0(B_0-B_1)-\alpha A_1(B_0+B_1).\nn
\end{align}
Then we can give two types of SOS decompositions for generalized tilted-CHSH operator in Eq. \eqref{eq:T2}. The first decomposition is given as
\begin{align}\label{SOS1}
\widehat{\mathcal{B}}_{[\alpha,\beta]}
=&\frac{1}{\Delta+2\eta_{[\alpha,\beta]}}\{
(\widehat{\mathcal{B}}_{[\alpha,\beta]})^2+\alpha^2(\beta A_1-S_0)^2\nn\\
&+(\alpha^2-1)
[(\xinhui{-}\beta A_0+\frac{\eta_{[\alpha,\beta]}}{\alpha^2+1}-A_1(B_0-B_1))^2\nn\\
&+(-\frac{\eta_{[\alpha,\beta]}\alpha}{\alpha^2+1}A_0+B_0+B_1)^2]\}.
\end{align}
And the second one is 
\begin{align}\label{SOS2}
\widehat{\mathcal{B}}_{[\alpha,\beta]}
=&\frac{\alpha^2}{\Delta+2 \eta_{[\alpha,\beta]}}
\big\{\frac{\alpha^2-1}{\alpha^2}
[(-\frac{\eta_{[\alpha,\beta]}\alpha}{\alpha^2+1}A_0+B_0+B_1)^2\nn\\
&+(\xinhui{-}\beta A_0+\frac{\eta_{[\alpha,\beta]}}{\alpha^2+1}-A_1(B_0-B_1))^2]\nn\\
&+\big(2A_0-\frac{\eta_{[\alpha,\beta]}}{2\alpha}(B_0+B_1)+\frac{\beta}{2}S_1)^2\nn\\
&+\frac{1}{\xinhui{\alpha^2}}(2A_1-\frac{\eta_{[\alpha,\beta]}}{2} (B_0-B_1)+\frac{\beta}{2}S_2)^2 \big\}
\end{align}
where \xinhui{$\Delta=2(\alpha^2-1)\sqrt{\frac{\beta^2+4}{\alpha^2+1}}$}.

For the special case $\alpha=1$ of standard tilted-CHSH inequality, our result gives the following decomposition:
\begin{align}
\widehat{\mathcal{B}}_{[1,\beta]}=\frac{1}{2\eta_{[1,\beta]}}[(\widehat{\mathcal{B}}_{[1,\beta]})^2+(\beta A_1-S_0)^2],
\end{align}
and
\begin{align}
\widehat{\mathcal{B}}_{[1,\beta]}=&\frac{1}{2\eta_{[1,\beta]}}
[(2A_0-\eta_{[1,\beta]}\frac{B_0+B_1}{2}+\frac{\beta}{2}S_1)^2\nn\\
&+(2A_1-\eta_{[1,\beta]}\frac{B_0-B_1}{2}+\frac{\beta}{2}S_2)^2], 
\end{align}
which reproduce the results in Ref. \cite{2015Bamps}. 
Thus we develop a family of SOS decompositions for generalized tilted-CHSH inequalities, which is beyond the  standard form.

\xinhui{If one observes the maximal quantum violation of the generalized tilted-CHSH inequality in Eq. \eqref{eq:T2} by any state $\ket{\psi}$ and measurements $A_x$, $B_y$ for $x,y\in\{0,1\}$,  then each square of polynomial functions in two SOS decompositions acting on $\ket{\psi}$ is equal to zero, i.e., $P_i\ket{\psi}=0$. Then we can obtain the anti-commutation relations for the measurements operators acting on the underlying state from the two SOS decompositions \eqref{SOS1}--\eqref{SOS2}  as following (the details refer to Appendix)
\begin{subequations}
\begin{align}
(Z_A-Z_B)\ket{\psi}&=0,\label{AP3:sosR1}\\
(\sin\theta X_A(\mathbb{I}+Z_B)-\cos\theta X_B(\mathbb{I}-Z_A))\ket{\psi}&=0.\label{AP3:sosR2}
\end{align}
\end{subequations}}

\xinhui{Next we will show that these algebraic relations lead to  self-testing statement for any partially entangled two-qubit state.}

\begin{figure}[ht!]
\centering
\includegraphics[scale=0.35]{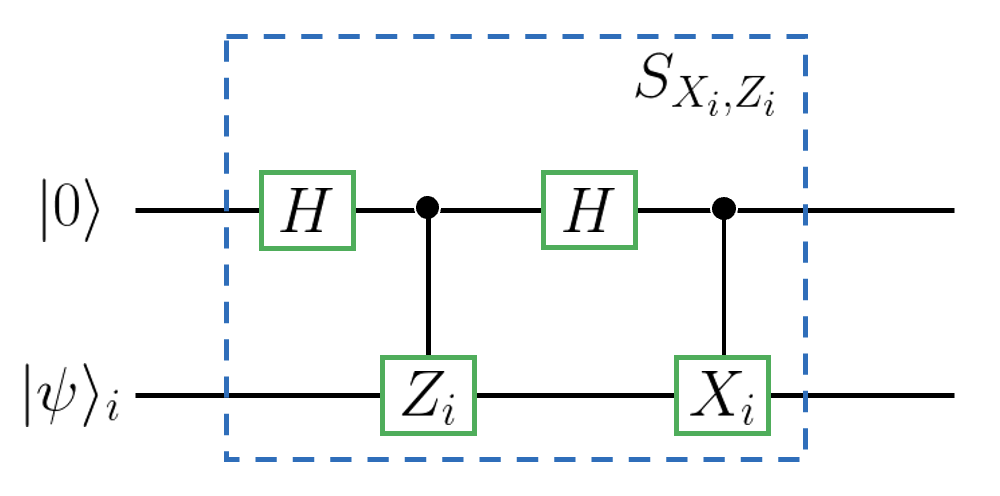}
\caption{The local unit $S_i$ of the swap gate $S=S_A \otimes S_B$ for $i\in\{A,B\}$. Each unit acts on the corresponding particle of the $\ket{\psi}$ and one ancillary qubit prepared in the state $\ket{0}$. $H$ is the standard Hadamard gate, $Z_i$ and $X_i$ are controlled by the auxiliary qubits.}\label{Fig:Swap}
\end{figure}

\textit{b. Self-testing of partially entangled states.}  \xinhui{Basing on the Definition \ref{Def1} of self-testing, one needs to construct the isometry map such that the underlying system can extract out the information of target state.  The isometry is a virtual protocol, all that must be done in laboratory is to query the boxes and derive
$p(a,b|x,y)$. The useful way is so-called swap method and  the isometry is shown in Fig. \ref{Fig:Swap}. The idea of swap method is from the ideal case. If the state $\ket{\psi}$ is indeed 
two-qubit and the operators are $Z=\sigma_z$, $X=\sigma_x$, the swap operations could extract the state $\ket{\psi}$ into ancilla system.  However, in the device-independent framework, it can not assume the dimension of the inner state or any form of the operators. Hence $Z$ and $X$ are constructed basing on real performed measurements $A_x$ and
$B_y$ such that one can swap out the desired states and measurements, 
as shown in Definition \ref{Def1}. } Therefore, we define the unitary operators of Alice and Bob as
\begin{align}
Z_A=A_0,\;\;Z_B&=\frac{B_0+B_1}{2\cos\mu}, \nonumber\\  
X_A=A_1,\;\;X_B&=\frac{B_0-B_1}{2\sin\mu}.
\end{align}
After this isometry, the underlying systems and the trusted auxiliary qubits will be 
\begin{align}\label{isometry}
\Phi(\ket{\psi})=&\frac{1}{4}[(\mathbb{I}+Z_A)(\mathbb{I}+Z_B)\ket{\psi}\ket{00}\nn\\
&+X_B(\mathbb{I}+Z_A)(\mathbb{I}-Z_B)\ket{\psi}\ket{01}\nn\\
&+X_A(\mathbb{I}-Z_A)(\mathbb{I}+Z_B)\ket{\psi}\ket{10}\nn\\
&+X_AX_B(\mathbb{I}-Z_A)(\mathbb{I}-Z_B)]\ket{\psi}\ket{11}.
\end{align}

 From relation \eqref{AP3:sosR1}, the second and third terms of Eq. \eqref{isometry} cancel to be zero. Then relation \eqref{AP3:sosR2} eventually leads Eq. \eqref{isometry} to be
\begin{align}
\Phi (\ket{\psi})=&\frac{\mathbb{I}+Z_A}{2}\ket{\psi}\ket{00}+\frac{\mathbb{I}+Z_A}{2}\frac{\sin\theta}{\cos\theta}\ket{\psi}\ket{11}\nn\\
=&\ket{junk}\otimes\ket{\overline{\psi}},
\end{align}
where
\begin{align}
\ket{junk}=\frac{\mathbb{I}+Z_A}{2\cos\theta}\ket{\psi}.
\end{align} 
Thus the underlying state is equal to the optimal target form  $\ket{\overline{\psi}}=\cos\theta\ket{00}+\sin\theta\ket{11}$ with $\sin2\theta=\sqrt{\frac{4-\alpha^2\beta^2}{4+\beta^2}}$. This completes the self-testing statement. 

\xinhui{The generalized tilted-CHSH operator $B_{[\alpha,\beta]}$ with two parameters such that the optimal measurements for one party could rotate on Pauli $x$-$z$ plane respect to the target one satisfying $\alpha\tan\mu={\sqrt{\frac{4-\alpha^2\beta^2}{4+\beta^2}}}$. It results in that in a self-testing scenario involving different  target $\ket{\overline{\psi}_i}$ defined as Eq. \eqref{eq:St2}, one can choose a common measurement settings \eqref{eq:M} satisfying $\tan\mu=\frac{\sin2\theta_i}{\alpha_i}$ to  construct the Bell inequality $\eta_{[\alpha_i,\beta_i]}$ for each target state. In turn, the maximal violations uniquely certify the family of states $\ket{\overline{\psi}_i}$. Thus, we complete the proof of  Theorem \ref{Theorem1}. }

To be specific, $\mathcal{B}_{[\alpha,\beta]}$ has two special forms when $\beta=0$ and $\alpha=1$, which corresponds to  biased CHSH \cite{Lawson2010} and standard tilted-CHSH operators \cite{Bancal}, respectively. 


--\textit{Biased CHSH inequality.} If $\beta=0$, the Bell inequality in Eq. \eqref{eq:T2} is 
simplified as a symmetrical biased CHSH operator 
\begin{align}
   \mathcal{B}_{[\alpha,0]}=\alpha(A_0B_0+A_0B_1)+A_1B_0-A_1B_1\leq2 \alpha,\label{eq:T0}
\end{align}
where $\alpha=\frac{1}{\tan\mu}$, which belongs to the whole set of self-testing  criteria for Bell state \cite{2016Wang}. Its maximal quantum violation $\eta_{[\alpha,0]}=\frac{4}{\sqrt{1+\tan\mu}}$ is able to self-test maximum entangled state $\ket{\overline{\psi}}=\frac{1}{\sqrt{2}}(\ket{00}+\ket{11})$ and measurements setting \eqref{eq:M}.


--\textit{Standard tilted-CHSH inequality.} If $\alpha=1$, the Bell  
inequality in Eq. \eqref{eq:T2}  turns to be the standard form
\begin{align}
  \mathcal{B}_{[1,\beta]}=\beta A_0+A_0B_0+A_0B_1+A_1B_0-A_1B_1\leq 2+\beta,\label{eq:T1}
\end{align}
with $\beta=2\sqrt{2\cos^2\mu-1}$. The maximal quantum violation of this inequality is given
by $\eta_{[1,\beta]}=\sqrt{8+2\beta^2}=4\cos\mu$, achievable with the measurement
settings in Eq. \eqref{eq:M}  and  satisfies $\sin2\theta=\sqrt{\frac{4-\beta^2}{4+\beta^2}}$.

\section{robustness analysis}\label{Sec:Rob}

\begin{figure}[ht]
\centering
\includegraphics[scale=0.35]{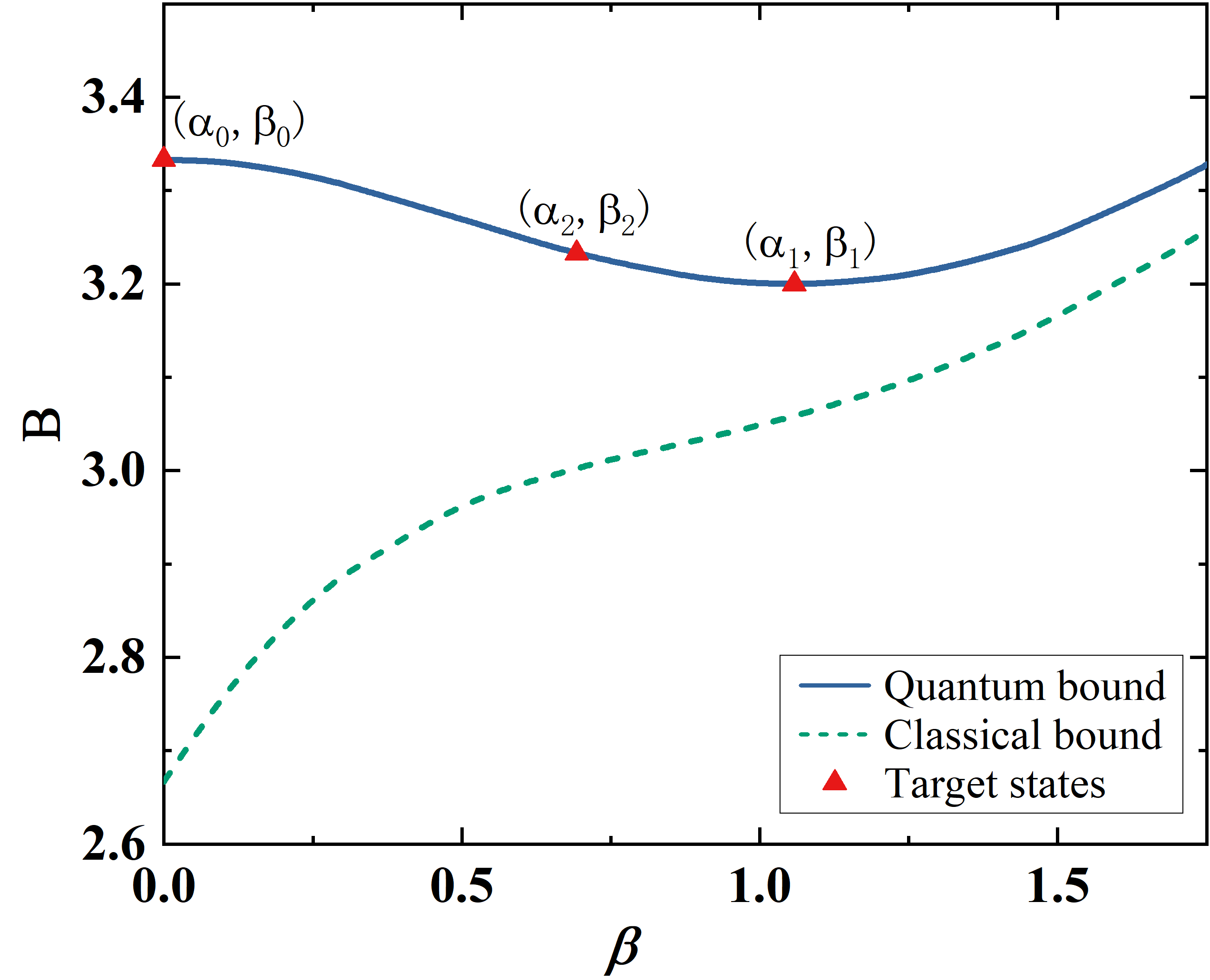}
\caption{The maximal violations of Bell inequalities $\mathcal{B}_{[\alpha,\beta]}$ with respect to $\beta$ for $\tan\mu=3/4$. 
The blue solid and green dotted lines represent the maximal quantum and classical bounds, respectively. The red triangles on blue line are obtained  with $[\alpha_i,\beta_i]$ for i=0,1,2 which are ideal scenarios to self-test the three target states in Eq. \eqref{Rst}.}\label{Fig:CQB}
\end{figure}

 \xinhui{If the observed statistics deviate from the ideal ones, one can estimate how far the actual state and measurements are from the ideal ones, a property known as robustness.} Here the robust self-testings of the different sources are analysed using numerical tool named Navascu\'{e}s-Pironio-Ac\'{i}n (NPA) hierarchy and SDP method \cite{NPA2008,Bancal}. \xinhui{For convenience of calculations,} we take $\mu=\arctan\frac{3}{4}$($\mu \approx 0.208\pi$) in measurements settings $\eqref{eq:M}$ as an example to self-test the following three sates 
\begin{align}\label{Rst}
\ket{\overline{\psi}_0}=&\frac{1}{\sqrt{2}}(\ket{00}+\ket{11}),\nn\\
\ket{\overline{\psi}_1}=&\cos\theta\ket{00}+\sin\theta\ket{11}\; \text {for } \theta=\frac{1}{2}\arcsin(\frac{3}{4}),\nn\\ %
\ket{\overline{\psi}_2}=&\frac{1}{2}(\ket{00}+\sqrt{3}\ket{11}),
\end{align}
which satisfy $\tan\mu=\frac{\sin2\theta_i}{\alpha_i}$ with $\alpha_0=\frac{4}{3}, \alpha_1=1$ and $\alpha_2=\frac{2}{\sqrt{3}}$. 
\xinhui{Choosing these three states not only associates with three special Bell operators (biased CHSH, standard tilted-CHSH and generalized tilted-CHSH operators), but also results in a convenience for the calculations in robustness analysis. To self-test these three target states, the parameter $\beta_i$ in Bell inequality \eqref{eq:T2} are set as $\beta_0=0$, $\beta_1=\frac{2\sqrt{7}}{5}$ and $\beta_2=\frac{2\sqrt{3}}{5}$ satisfying $\sin2\theta_i=\sqrt{\frac{4-\alpha_i^2\beta_i^2}{4+\beta_i^2}}$, respectively. By substituting $[\alpha_i,\beta_i]$ into  $\mathcal{B}_{[\alpha_i,\beta_i]}$ in Eq.\eqref{eq:T2} for $i=0,1,2$, it can be found the first two pairs respectively recover to biased inequality \eqref{eq:T0}  and tilted-CHSH inequalities  \eqref{eq:T1},  while the third pair is  complex.} \xinhui{Particularly, by fixing $\tan\mu=\frac{3}{4}$, the parameter $\alpha$ can be express by $\beta$, i.e., $\alpha=\frac{8}{\sqrt{25\beta^2+36}}$, thus we can  plot the bounds of Bell inequality $\mathcal{B}_{[\alpha_i,\beta_i]}$ respect to $\beta$.} 
 \xinhui{As shown as in Fig. \ref{Fig:CQB}, with the increasing of $\beta$,
the classical bound tends nearly but no more than the quantum.  The maximal quantum bounds for the ideal self-testing of the three states \eqref{Rst} and measurement settings \eqref{eq:M} are presented by red triangles in Fig. \ref{Fig:CQB}. The gap between classical and quantum bounds at $(\alpha_0,\beta_0)$ is much larger than the other points.}

After the isometry given in Fig. \ref{Fig:Swap}, the trusted auxiliary systems will be left in the state
\begin{align}
    \rho_{swap}=\sum_{ijst}C_{ijst}\ket{j}\bra{i}\otimes\ket{t}\bra{s},
\end{align}
where $C_{ijst}=\frac{1}{16}tr_{AB}[(1+Z_A)^{1-i}(X_A-Z_AX_A)^i(1+Z_A)^{1-j}(X_A-X_AZ_A)^j\otimes 
    (1+Z_B)^{1-s}(X_B-Z_BX_B)^s(1+Z_B)^{1-t}(X_B-X_AZ_B)^t]$. 
Finally, we shall be able to express the fidelity for $i=0,1,2$
\begin{align}\label{eq:SDPobj}
f_i = \bra{\overline{\psi}_i} \rho_{swap} \ket{\overline{\psi}_i}.
\end{align}
Here $f_i$ is a linear function of two types of operator expectations: some observed behavior and some non-observable correlations which 
involve different measurements on the same party, 
such as $\bra{\psi}M^a_x M^{a'}_{x'}\ket{\psi}$ with $x\neq x'$ which are left as variables. 

To get a lower bound on the fidelity, one
needs to minimize the fidelity running over all the states and measurements satisfying 
observed statistics. Optimizations over the set of quantum momenta are
computationally hard, specially for the underlying Hilbert space dimension is unknown. To resolve this technical difficulty, here we employ the NPA hierarchy which was introduced in Refs. \cite{NPA2008,Bancal} to bound fidelity. \xinhui{The NPA hierarchy works as follows. Consider a generic state and measurement operators $\{\ket{\psi},A_x,B_y\}$.  Then, define sets $Q_l$ (each corresponding to a level of the hierarchy comprised of the identity operator and all (non-commuting) products of operators $A_x$, $B_y$ up to to degree $l$, e.g. $Q_1=\{\mathbb{I},A_x,B_y\}$, $Q_2=\{Q_1\}\cup\{Q_1^{i}Q_1^{j}\}$,...,$Q_k=Q_k\cup\{Q_k^{i}Q_1^{j}\}$, where $Q_{k}^i$ is the $i\text{th}$ element of $Q_k$. Define the moment matrix of order $l$, $\Gamma^{k}$ by $\Gamma_{ij}=\bra{\psi}{Q^i_l}^\dagger Q^j_l\ket{\psi}$. For any
state and measurements $\{\ket{\psi},A_x,B_y\}$, the matrix $\Gamma^l$
is Hermitian positive semidefinite and satisfies some linear constraints given by the orthogonality conditions of the measurement operators \cite{NPA2008}. 
Thus we can tackle the optimization problem by minimizing the corresponding elements of the matrix $\Gamma$ under linear constraints on $\Gamma\geq0$ to obtain certified lower bounds to the optimal solution
\begin{align}\label{eq:MSDP}
\min \quad & f_i = \bra{\overline{\psi}_i} \rho_{swap} \ket{\overline{\psi}_i} \nn\\
\text{s.t.}\quad  &\Gamma \geq 0, \\
& \mathcal{B}_{[\alpha_i,\beta_i]}=\text{observed violation value} \;\;i=0,1,2\nn
\end{align}
where $\Gamma$ is a $46 \times 46$ moment matrix of quantum local level
one $Q_1=\{I,Z_A,X_A\}\otimes\{I,Z_B,X_B\}$ and augmented by necessary terms such as $Z_AX_A$, $X_AZ_BX_B$, $Z_AX_AZ_B$, et.al., to express the fidelity.  Thus we are able to formulate this problem as a SDP, a type of convex optimization for which there exist efficient numerical solvers to find global minima and which also return the error bounds on the optimal guess.}

\begin{figure}[ht]
\centering
\includegraphics[scale=0.36]{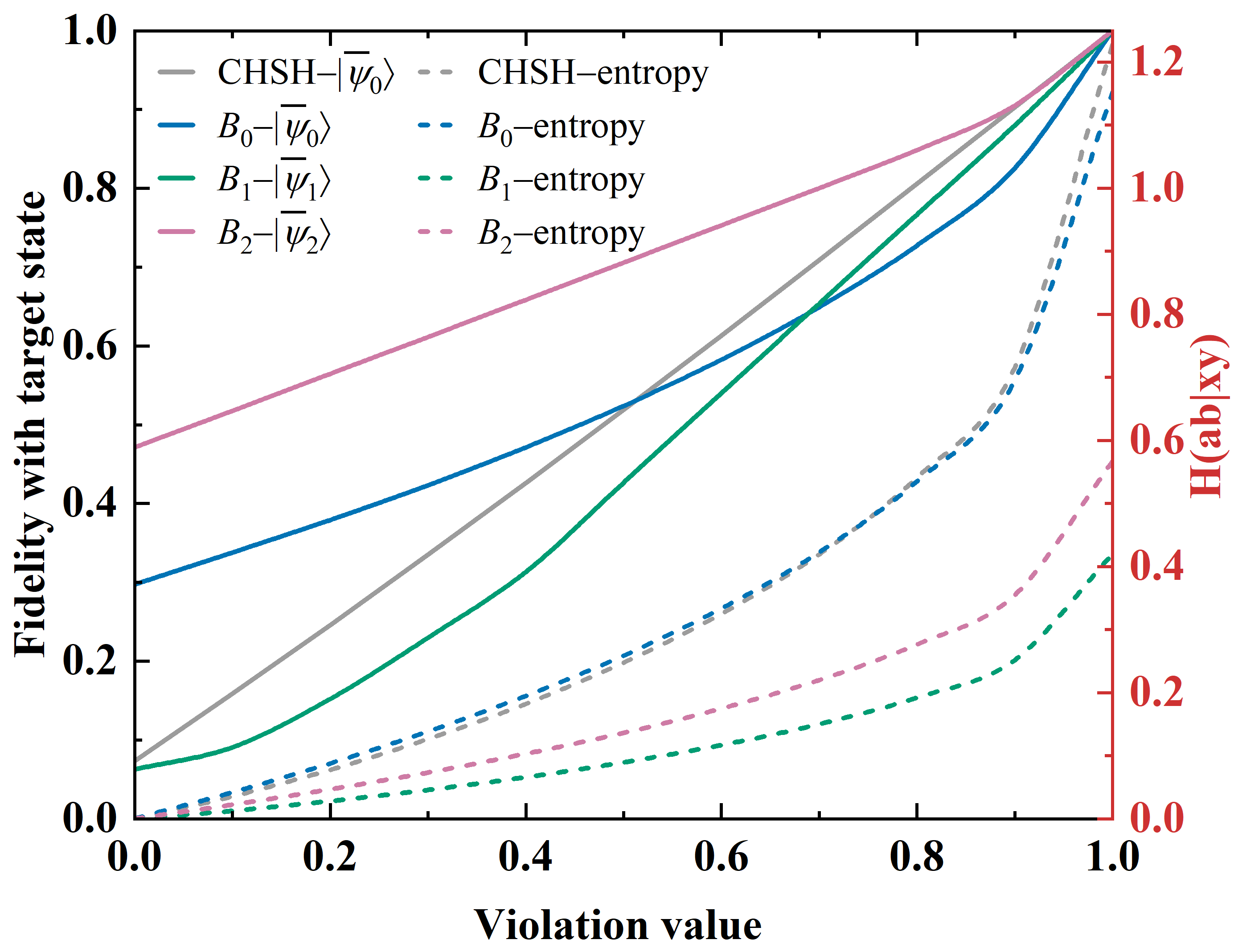}
\caption{The lower bounds of fidelity with target state (left vertical axis) and randomness entropy (right vertical axis) with respect to the observation of Bell inequality. We denote $B_{[\alpha_i,\beta_i]}$ as $B_i$ for $i=0,1,2$.
The observed violation is made a transform $V_i=\frac{obs-C_{[\alpha_i,\beta_i]}}
{\eta_{[\alpha_i,\beta_i]}-C_{[\alpha_i,\beta_i]}}$. 
The fidelity with Bell state $\ket{\overline{\psi}_0}$ using standard CHSH and biased form are presented in gray and blue purple solid lines, respectively.  The fidelity with $\ket{\overline{\psi}_1}$ and $\ket{\overline{\psi}_2}$ are represented in green and purple solid lines, respectively.
The randomness for each Bell inequality is plotted with right vertical axis in dash color lines. }\label{Fig:RobQRNG}
\end{figure}

The robustness analyses are shown in Fig. \ref{Fig:RobQRNG} with left vertical axis. For the Bell state, the fidelity bound by standard CHSH is higher than using biased one when the violations close to the maximal quantum bounds. This result agrees on the work in Ref. \cite{2016Wang} that for $\mu$ closer to $\frac{\pi}{4}$, the criterion has a better capacity in noise tolerance. For partial entangled states, the tilted-CHSH inequality is fragile  to noise for weakly entangled state, while the generalized tilted-CHSH operator has better performance. \xinhui{This may result from that, for $\mu=\arctan(\frac{3}{4})$, the gap between the quantum and classical bounds for generalized tilted-CHSH operator is larger than the standard one shown in Fig. \ref{Fig:CQB}, thus provides better distinguishability between different states.} 

\xinhui{So far, we have provided the scheme that self-test  different entangled states by fixed measurements settings based on generalized tilted-CHSH inequality, which is robust in tolerance of noise. Next, we demonstrate the applications of our results to simplify the implementations of secure quantum information tasks such as quantum key distribution (QKD), quantum random number generation (QRNG) and quantum private query (QPQ).}
\section{Applications in quantum information tasks}\label{sec:Apl}

Quantum systems with self-testing property play important roles in quantum information processing. Especially, for the protocols which require a high demand for the security, self-testing is able to guarantee the security independently with the devices.
This is precisely the fact that motivates the device independent (DI) quantum information processing. 
In last few years, the DI technologies 
 have been studied  intensively. 
 Among of them, QKD,  QRNG and QPQ 
 as the core and bases of quantum cryptography, have gained huge attentions. 

\subsection{Device independent quantum key distribution and private query}

DIQKD allows distant parties to create and share a
cryptographic key, whose security relies
only on the certification of nonlocal quantum correlations \cite{DIQKD2014}. In the simplest protocol, 
entangled particles  are repeatedly prepared and distributed between two parties, Alice and Bob.
Alice holds two measurements $A_x$ for $x\in\{0,1\}$, Bob has three measurements $B_y$, $y\in\{0,1,2\}$. To ensure
the security of the task, Alice and Bob perform CHSH test by randomly choosing two measurements $A_x$ and $B_y$, $x,y\in\{0,1\}$ respectively to certify the source device independently. The maximal quantum bound $2\sqrt{2}$ implies that the source is the maximum entangled state $\ket{\overline{\psi}}=\frac{1}{\sqrt{2}}(\ket{00}+\ket{11})$ and measurements are 
$A_0=\sigma_z, A_1=\sigma_x$, $B_0=\frac{\sigma_x+\sigma_z}{\sqrt{2}}$ and $B_1=\frac{\sigma_x-\sigma_z}{\sqrt{2}}$. Then the measurement $A_0$ for Alice and the last one  $B_2=\sigma_z$ for Bob are used to extract secure key.

Later motivated by this idea, a DIQPQ protocol was proposed in Ref. \cite{Maitra2017}. 
In the protocol, 
Alice and Bob share an entangled state with $\frac{1}{\sqrt{2}}(\ket{0}_A\ket{\phi_0}_B+\ket{1}_A\ket{\phi_1}_B)$ where
\begin{align}
    \ket{\phi_0}_B=\cos\frac{\theta}{2}\ket{0}+\sin\frac{\theta}{2}\ket{1},\nn\\
    \ket{\phi_1}_B=\cos\frac{\theta}{2}\ket{0}-\sin\frac{\theta}{2}\ket{1}.
\end{align}
Before the process of QPQ \cite{Yang2014}, Alice and Bob perform a CHSH-like test to certify the source and measurements, which 
guarantee the measurements for Alice are in the basis $\{\ket{0},\ket{1}\}$ and $\{\ket{+},\ket{-}\}$, 
and Bob's are in basis $\{\ket{\psi_0},\ket{\psi_0^\perp}\}$ or $\{\ket{\psi_1},\ket{\psi_1^\perp}\}$. If the outcome for Bob is 
$\ket{\psi_0^\perp}(\ket{\psi_1^\perp})$, he can conclude that the raw key bit at Alice must be 0(1). 
Bob and Alice execute classical post processing, so that information of Bob on the key
reduces to one bit or more. Alice knows the whole key, whereas Bob generally knows several bits of the key.

According to our work, the measurements for Alice and Bob in above two protocols can be set as
\begin{align}
    A_0&=\sigma_z \;B_0=\cos2\theta \sigma_z+\sin2\theta\sigma_x,\;\nn\\
    A_1&=\sigma_x \;B_1=\cos2\theta \sigma_z-\sin2\theta\sigma_x,\;
\end{align}
which are available to self-test the two entangle sources in DIQKD and DIQPQ tasks at same time. For instance, the parameters in Eq. \eqref{eq:T2} can be set according to the task, i.e., $\alpha=\frac{1}{\tan2\theta}$ and $\beta=0$ for DIQKD protocol,  $\alpha=1$ and $\beta=\frac{2\cos\theta}{\sqrt{1+\sin^2\theta}}$ for DIQPQ protocol. In this way, different entangled states can not only be certified as the source in QKD, but also be used to generate secure key in QPQ task. This simplifies the  measurement resources to achieve different quantum information processing.

\subsection{Device independent quantum random number generation}\label{Sec:QRNG}
DIQRNG is able to access randomness by observing the violation of Bell inequalities without any assumptions on the source and measurement device. The randomness of the output pairs conditioned on the input pairs for the entangled pairs can be quantified by the min-entropy \cite{DIQRNG2010} $H_\infty(ab|xy)=-\log_2\max p(ab|xy)$. For a given observed violation $V_i$ of Bell inequality $B_{[\alpha_i,\beta_i]}$, where $V_i$ is defined in the caption of Fig. \ref{Fig:RobQRNG}, we are able 
to obtain a lower bound on the min-entropy
\begin{align}
    H_\infty(ab|xy)\geq V_i
\end{align}
satisfied by all quantum realizations of the Bell scenario. Let $P^*(ab|xy)$ denote the solution to the following 
optimization problem
\begin{align}
\text{obj} \quad &P^*(ab|xy)=\max p(ab|xy)\nn\\
\text{s.t.}\quad  &\Gamma \geq 0, \nn\\
& \mathcal{B}_{[\alpha_i,\beta_i]}=\text{observed violation value} \;\;i=0,1,2\nn
\end{align}
where the optimization is carried over all states $\rho$ and all measurement operators, defined over Hilbert spaces of arbitrary dimension. The minimal value of the min-entropy compatible with the Bell violation $V$ and quantum theory is then given by $H_\infty(ab|xy) =-\log_2 \max_{ab}P(ab|xy)$.

Using the same Bell inequalities in Sec. \ref{Sec:Rob} and running SDP with NPA  hierarchy,  
we plot the lower bounds of the entropy as shown in Fig. \ref{Fig:RobQRNG} with right vertical axis. 
It is noteworthy to point that in a device independent framework, if the violation is maximal quantum bound for CHSH $2\sqrt{2}$, 
the randomness is obtained with 1.2283 bits and the underlying structure is Bell state with orthogonal basis \cite{DIQRNG2010}. Here, we shown that if the two observers self-test the state with the biased basis as Eq. \eqref{eq:M}, the lower bound of randomness is 1.1519 bits, which is slightly lower than the CHSH scenario. However, we point out with this biased basis, the random numbers can also be certified from partially entangled states. As to the two partially entangled target states whose concurrence are respective $C(\ket{\overline{\psi}_1})=\frac{3}{4}$ and  $C(\ket{\overline{\psi}_2})=\frac{\sqrt{3}}{2}\approx0.866$, the secure randomness can be extracted with 0.4195 bits and 0.5669 bits, respectively. In other words, with this measurement settings, we can extract randomness in both the maximum entangled and  partially entangled states.

\section{Conclusion}\label{sec:Conclusion}

Self-testing results are usually known for a set of quantum state and corresponding measurement simultaneously. However different entangled resources are needed for various quantum information tasks with diverse requirements. In this paper we proposed a scheme that self-tests a family of entangled states with different entanglement degree by the same fixed measurement settings. By providing the SOS decompositions of generalized tilted-CHSH inequality, we extend the self-testing criteria of general  two-qubit states with two binary measurements per party. The previous work based on symmetric biased CHSH \cite{Lawson2010} and standard tilted-CHSH operators \cite{2015Bamps} can be regarded as special cases of this criteria.The self-testing criteria obtained in our work are appealing from two aspects. For general two-qubit entangled states,  we broaden its self-testing criteria. The self-testing can be carried out with a series of different measurements settings on Pauli $x$-$z$ plane, by setting different values of $\alpha$ in generalized tilted-CHSH inequality. More importantly, different entangled states can be self-tested by maximally violating corresponding Bell inequalities with the same fixed measurement settings. This can  simplify the measurement instruments of self-testing in experimental realization. Moreover our scheme demonstrate satisfactory robustness in tolerance of noise.


Furthermore, our scheme can provide secure certification for different
device-independent quantum information processing tasks with fewer resource.  
This work is instrumental to improve the practical performance of self-testing. In addition, this work is of intrinsic interest to the foundational studies on Bell nonlocality and quantum certification. 
In the future, it is interesting to study more Bell nonlocalities with self-testing property and find more criteria with the same measurements for different states. 


\section*{Acknowledgments}
This work was supported by National Natural Science Foundation of China (Grant No. 62101600, 51890861, 11974178, 62201252), China University of Petroleum Beijing (Grant No.ZX20210019), State Key Laboratory of Cryptography Science and Technology (Grant No. MMKFKT202109), National Key Research and Development Program of China
(2019YFA0705000) and Leading-edge technology Program of Jiangsu Natural Science Foundation (No. BK20192001).



\section*{Appendix. The SOS decomposition for generalized tilted-CHSH inequality}\label{app}

\setcounter{equation}{0}
\renewcommand{\theequation}{A.\arabic{equation}}
\renewcommand{\thesubsection}{A.\arabic{subsection}}

We provide the way to obtain the  SOS decompositions of generalized tilted-CHSH operator \eqref{Ap2:til} 
\begin{align}\label{Ap2:til}
\mathcal{B}_{[\alpha,\beta]}=\beta A_0+\alpha A_0B_0+\alpha A_0B_1+A_1B_0-A_1B_1
\end{align}
where $\alpha\geq 1$, in detail.

The optimal quantum violation of \eqref{Ap2:til} is proved to be $\eta_{[\alpha,\beta]}=\sqrt{(1+\alpha^2)(4+\beta^2)}$ by optimizing over all quantum states and measurements \cite{Acin2012}. 
The bound implies the operator 
$\widehat{\mathcal{B}}=\eta_{[\alpha,\beta]}\mathbb{I}-\mathcal{B}_{[\alpha,\beta]}$ 
is positive semidefinite for all possible quantum states and measurement
operators $A_x$ and $B_y$. This in turn can be proven by providing a set of operators $\{P_i\}$
which are polynomial functions of $A_x$ and $B_y$
such that
\begin{align}\label{eq:SOS}
\widehat{\mathcal{B}}_{[\alpha,\beta]}=\sum_i P^\dagger_i P_i
\end{align}
holds for any set of measurement operators satisfying the
algebraic properties $A^2_x=\mathbb{I}$, $B^2_y=\mathbb{I}$ and
$[A_x,B_y]=0$. The form \eqref{eq:SOS} is called a
SOS decomposition.

Our goal is to find SOS decompositions of generalized tilted-CHSH inequalities as in Eq. \eqref{eq:SOS} in terms of
a set of polynomials $\{P_i\}$. Our techniques on SOS decompositions is based on Ref. \cite{2015Bamps}.
For simplicity, the search space
is restricted to the span of a canonical basis of nine monomials
\begin{align}
\mathcal{S}_{1+AB}=\{\mathbb{I},A_0,A_1\}\otimes\{\mathbb{I},B_0,B_1\}.
\end{align}

Let $\{R_i\}_i$ denote the different bases of the vector space of polynomial $P_i$. So
$P_i$ can be expressed by bases $P_i=\sum_\mu q^\mu_i R_\mu$. 
The $\widehat{\mathcal{B}}$ is rewritten as
\begin{align}\label{eq:SOSM}
\widehat{\mathcal{B}}=\sum_{\mu v}\sum_{i} R^\dagger_\mu q^\mu_i q^v_iR_v=\sum_{\mu v} R^\dagger_\mu M^{\mu v}R_v.
\end{align}

The task turns to find a positive semidefinite matrix
$M$ such that Eq. \eqref{eq:SOSM} holds. By decomposing both sides
of the equality $\widehat{\mathcal{B}}=\sum_{\mu v}M^{\mu v}R^\dagger_\mu R_v$
in a basis of the quadratic products of all elements in $S_{1+AB}$, 
we obtain a canonical basis with size 25 for these products as
\begin{align}
S^2_{1+AB}=&\{\mathbb{I},A_0,A_1,A_0A_1,A_1A_0\}\nn\\
&\otimes\{\mathbb{I},B_0,B_1,B_0B_1,B_1B_0\}.
\end{align}
Writing $R^\dagger_\mu R_v=F^i_{\mu v}E_i$,  where $E_i$ takes over 
$S^2_{1+AB}$ and each $F_i$ is a matrix of coefficients such that
$\widehat{\mathcal{B}}=s^iE_i$. Then the SOS condition reduces to
\begin{align}
s^i=Tr(M^\dagger F_i) \text{ for }i=1,2,...,25.
\end{align}
The left thing is to solve a set of 25 linear equality constraints
on $M$ as well as the positive semidefiniteness
constraint $M\geq 0$.

A valid SOS decomposition
for $\widehat{\mathcal{B}}_{[\alpha,\beta]}$ must be made up of terms for which $P_i(\cos\theta\ket{00}+\sin\theta\ket{11})$ vanishes
in this maximally violating quantum system. Indeed, writing the most
general $P$ in the search space as $r\cdot\mathbf{V}$ where
\begin{align}
\mathbf{V}=(\mathbb{I},A_0,A_1,B_0,B_1,A_0B_0,A_0B_1,A_1B_0,A_1B_1),
\end{align}
and demand the four components of $P_i\ket{\psi}$ vanish, Ref. \cite{2015Bamps} has shown that for $\alpha=1$
the space of candidates $P_i$ is spanned by the following five operators:
\begin{align}
-Z_A+Z_B\nn\\
-\mathbb{I}+Z_AZ_B\nn\\
-\mathbb{I}+c Z_B+X_AX_B\nn\\
-X_A+sX_B+cX_AZ_B\nn\\
-cX_A+sZ_AX_B+X_AZ_B
\end{align}
where $c=\cos2\theta$, $s=\sin2\theta$ and operators $Z$ and $X$ are
defined as
\begin{align}\label{Ap2:meas}
 Z_A=A_0, \;&Z_B=\frac{B_0+B_1}{2\cos\mu},\nn\\
X_A=A_1,\;&X_B=\frac{B_0-B_1}{2\sin\mu} .  
\end{align}

We express the maximal violation of the generalized tilted-CHSH operator \eqref{Ap2:til} with $\alpha$ and $\theta$
\begin{align}
\eta_{[\alpha,\theta]}=\frac{2(\alpha^2+1)}{\sqrt{\alpha^2+\sin^22\theta}}
\end{align}
which only be achieved by the state $\ket{\psi}=\cos\theta\ket{00}+\sin\theta\ket{11}$ and corresponding measurements $A_x$ and $B_y$ for $x,y\in\{0,1\}$ defined as \eqref{Ap2:meas} and parameters satisfy
\begin{align}
\beta=\frac{2\cos2\theta}{\sqrt{\alpha^2+\sin^22\theta}},\nn\\
\cos\mu=\frac{\alpha}{\sqrt{\alpha^2+\sin^22\theta}},\\
\sin\mu=\frac{\sin2\theta}{\sqrt{\alpha^2+\sin^22\theta}}.\nn
\end{align}

Now we choose the basis ${R_i}=\{r_i\cdot \mathbf{V}\}$ 
for the subspace containing the SOS polynomials for generalized tilted-CHSH \eqref{Ap2:til}, 
and label the columns with the operators defining $\mathbf{V}$,
where the $r_i$ vectors are defined as such
\begin{widetext}
\begin{equation}
\begin{array}{ccccccccccc}
      & \mathbb{I} & A_0& A_1& B_0& B_1 &A_0B_0&A_0B_1&A_1B_0&A_1B1&\\
r_1=( &0  &-\frac{2\alpha}{\sqrt{\alpha^2+s^2}}& 0 &1 &1& 0 &0 &0 &0&)\\
r_2=( &-\frac{2\alpha}{\sqrt{\alpha^2+s^2}} &0& 0 &0 &0& 1 &1 &0 &0&)\\
r_3=( &-\frac{2}{\sqrt{\alpha^2+s^2}}  &0& 0 &\frac{c}{\alpha} &\frac{c}{\alpha}& 0 &0 &1 &-1&)\\
r_4=( &0  &0& -\frac{2}{\sqrt{\alpha^2+s^2}} &1 &-1& 0 &0 &\frac{c}{\alpha} &\frac{c}{\alpha}&)\\
r_5=( &0  &0& -\frac{2c}{\sqrt{\alpha^2+s^2}} &0 &0& 1 &-1 &\frac{1}{\alpha} &\frac{1}{\alpha}&)\\
  \end{array}
\end{equation}
\end{widetext}
These basis operators separate the space in two isotypical subspaces, i.e., subspaces that fall
under the same irreducible representation of the cyclic
group: $R_{1,2,3}$ are invariant under the symmetry transformation
of $\mathcal{B}_{[\alpha,\beta]}$, while
$R_{4,5}$ change sign. 
The block structure of symmetric SOS matrices is therefore $3\oplus 2$, where the
first block corresponds to the trivial representation and
the second to the parity representation where the group
generator is represented by $-1$.

\begin{widetext}

For convenience, we define three classes CHSH operators:
\begin{align}
S_0&= A_0(B_0-B_1)+\frac{1}{\alpha}A_1(B_0+B_1),\nn\\
S_1&=\frac{1}{\alpha}A_0(B_0+B_1)-A_1(B_0-B_1),\\
S_2&=A_0(B_0-B_1)-\alpha A_1(B_0+B_1).\nn
\end{align}
Then  we can provide two different SOS decompositions of the generalized CHSH-inequalities $\mathcal{B}_{[\alpha,\beta]}$ in Eq. \eqref{Ap2:til}. The first one can be given as
\begin{align}\label{AP2:SOS1main}
\widehat{\mathcal{B}}_{[\alpha,\beta]}
=&\frac{1}{\Delta+2\eta_{[\alpha,\beta]}}\big\{
(\alpha^2-1)\big((\frac{\cos2\theta}{\alpha} R_1-R_3)^2+R_1^2\big)
+\big(\frac{\cos2\theta}{\alpha}R_1-\alpha R_2-R_3\big)^2+\alpha^2R_5^2\big\}\nn\\
=&\frac{1}{\Delta+2\eta_{[\alpha,\beta]}}\big\{(\alpha^2-1)
[(-\beta A_0+\frac{\eta_{[\alpha,\beta]}}{\alpha^2+1}-A_1(B_0-B_1))^2\nn\\
&+(-\frac{\eta_{[\alpha,\beta]}\alpha}{\alpha^2+1}A_0+B_0+B_1)^2]+\big(\eta_{[\alpha,\beta]}
-\beta A_0-\alpha A_0(B_0+B_1)-A_1(B_0-B_1)\big)^2+\alpha^2R_5^2\big\}\nn\\
=&\frac{1}{\Delta+2\eta_{[\alpha,\beta]}}\{
(\alpha^2-1)
[(\xinhui{-}\beta A_0+\frac{\eta_{[\alpha,\beta]}}{\alpha^2+1}-A_1(B_0-B_1))^2
+(-\frac{\alpha\eta_{[\alpha,\beta]}}{\alpha^2+1}A_0+B_0+B_1)^2]\nn\\
&+\widehat{\mathcal{B}}_{[\alpha,\beta]}^2+\alpha^2(\beta A_1-(A_0(B_0-B_1)+\frac{1}{\alpha}A_1(B_0+B_1)))^2\}\nn\\
=&\frac{1}{\Delta+2\eta_{[\alpha,\beta]}}\{(\alpha^2-1)
[(\xinhui{-}\beta A_0+\frac{\eta_{[\alpha,\beta]}}{\alpha^2+1}-A_1(B_0-B_1))^2+(-\frac{\eta_{[\alpha,\beta]}\alpha}{\alpha^2+1}A_0+B_0+B_1)^2]\nn\\
&+\widehat{\mathcal{B}}_{[\alpha,\beta]}^2+\alpha^2(\beta A_1-S_0)^2\}
\end{align}
where $\Delta=2(\alpha^2-1)\sqrt{\frac{\beta^2+4}{\alpha^2+1}}$. 
The second decomposition is given as
\begin{align}\label{AP2:SOS2main}
\widehat{\mathcal{B}}_{[\alpha,\beta]}
=&\frac{\alpha^2}{\Delta+2\eta_{[\alpha,\beta]}}\big\{
\frac{\alpha^2-1}{\alpha^2}[(\frac{\cos2\theta}{\alpha} R_1-R_3)^2+R_1^2]
+\big(\frac{2(\alpha^2+1)}{\alpha \eta_{[\alpha,\beta]} }R_1
-\frac{\beta}{2}(\frac{1}{\alpha}R_2-R3)\big)^2+\frac{1}{\xinhui{\alpha^2}}\big(\frac{\eta_{[\alpha,\beta]}}{2}R_4-\frac{\beta}{2}R_5\big)^2\big\}\nn\\
=&\frac{\alpha^2}{\Delta+2 \eta_{[\alpha,\beta]}}
\big\{\frac{\alpha^2-1}{\alpha^2}
[(\xinhui{-}\beta A_0+\frac{\eta_{[\alpha,\beta]}}{\alpha^2+1}-A_1(B_0-B_1))^2
+(-\frac{\alpha\eta_{[\alpha,\beta]}}{\alpha^2+1}A_0+B_0+B_1)^2]\nn\\
&+\big(2A_0-\frac{\eta_{[\alpha,\beta]}}{2\alpha}(B_0+B_1)+\frac{\beta}{2}(\frac{1}{\alpha}A_0(B_0+B_1)-A_1(B_0-B_1)))^2\nn\\
&+\frac{1}{\xinhui{\alpha^2}}(2A_1-\frac{\eta_{[\alpha,\beta]}}{2} (B_0-B_1)+\frac{\beta}{2}(A_0(B_0-B_1)-\alpha A_1(B_0+B_1)))^2 \big\}\nn\\
=&\frac{\alpha^2}{\Delta+2 \eta_{[\alpha,\beta]}}
\big\{\frac{\alpha^2-1}{\alpha^2}
[(-\frac{\eta_{[\alpha,\beta]}\alpha}{\alpha^2+1}A_0+B_0+B_1)^2+(\xinhui{-}\beta A_0+\frac{\eta_{[\alpha,\beta]}}{\alpha^2+1}-A_1(B_0-B_1))^2]\nn\\
&+\big(2A_0-\frac{\eta_{[\alpha,\beta]}}{2\alpha}(B_0+B_1)+\frac{\beta}{2}S_1)^2+\frac{1}{\xinhui{\alpha^2}}(2A_1-\frac{\eta_{[\alpha,\beta]}}{2} (B_0-B_1)+\frac{\beta}{2}S_2)^2 \big\}
\end{align}
here $\Delta$ is defined as same in \eqref{AP2:SOS1main}

\end{widetext}

\xinhui{Hence, we complete the SOS  decompositions for $\widehat{\mathcal{B}}_{[\alpha,\beta]}$ such that $\widehat{\mathcal{B}}_{[\alpha,\beta]}=\sum_i P^\dagger_iP_i$ and $P_i$ are the polynomial functions of $A_x$ and $B_y$. The existence of SOS decompositions \eqref{AP2:SOS1main} and \eqref{AP2:SOS2main} 
 implies that any state $\ket{\psi}$ and operators $A_x$ and $B_y$ achieving the maximal quantum bound $\eta_{[\alpha,\beta]}$ will result in  $P_i\ket{\psi}=0$. Particularly, we are interested in the following four terms
\begin{subequations}
\begin{align}
 P1\ket{\psi}&=\widehat{\mathcal{B}}_{[\alpha,\beta]}\ket{\psi}\label{P1}\\
 P2\ket{\psi}&=(\beta A_1-S_0)\ket{\psi}\label{P2}\\
 P3\ket{\psi}&=(2A_0-\frac{\eta_{[\alpha,\beta]}}{2\alpha}(B_0+B_1)+\frac{\beta}{2}S_1) \ket{\psi}\label{P3}\\
 P4\ket{\psi}&=(2A_1-\frac{\eta_{[\alpha,\beta]}}{2} (B_0-B_1)+\frac{\beta}{2}S_2)\ket{\psi}\label{P4}
 \end{align}
\end{subequations}
which can be linearly combined to form the following operators 
\begin{subequations}
\begin{align}
(Z_A-Z_B)\ket{\psi}&=0,\label{R11}\\
(\sin\theta X_A(\mathbb{I}+Z_B)-\cos\theta X_B(\mathbb{I}-Z_A))\ket{\psi}&=0,\label{R22}
\end{align}
\end{subequations}
in the case of yielding the maximal quantum bound.}
\xinhui{The algebraic relations \eqref{R11}--\eqref{R22} established by the SOS decompositions \eqref{AP2:SOS1main}--\eqref{AP2:SOS2main} are necessarily satisfied by any quantum state and observables achieving the maximal quantum bound. Moreover, they are important for the self-testing of partially entangled states $\ket{\psi}=\cos\theta\ket{00}+\sin\theta\ket{11}$ using the isometry circuit.}











\bibliographystyle{unsrtnat}


\end{document}